\documentclass[aps,twocolumn,showpacs,amsmath,amssymb,floatfix,superscriptaddress]{revtex4-1}
\pdfoutput=1
\usepackage{graphicx}
\usepackage{amsmath}
\usepackage{amsfonts}
\usepackage{amssymb}
\usepackage{bm}
\usepackage{hyperref}
\usepackage{txfonts}

\usepackage{tikz}
\usepackage{circuitikz}
\usetikzlibrary{arrows}
\usetikzlibrary{calc,decorations.markings}

\newcommand{\meann}[1]{\langle #1 \rangle}

\newcommand{\beq}{\begin{equation}}
\newcommand{\eeq}{\end{equation}}

\newcommand{\rom}[1]{\uppercase\expandafter{\romannumeral #1\relax}}

\newcommand{\aff}{\affiliation{Department of Physics and Materials Science,
University of Luxembourg, L-1511 Luxembourg, Luxembourg}}

\begin{document}

\author{Sadeq S. Kadijani}
\author{Thomas L. Schmidt}
\author{Massimiliano Esposito}
\author{Nahuel Freitas}
\aff

\title{Heat transport in overdamped quantum systems}

\date{\today}

\begin{abstract}
We obtain an analytical expression for the heat current between two overdamped quantum oscillators interacting with local thermal baths at different temperatures. The total heat current is split into classical and quantum contributions. We show how to evaluate both contributions by taking advantage of the time scale separation associated with the overdamped regime, and without assuming the usual weak coupling and Markovian approximations. We find that non-trivial quantum corrections survive even when the temperatures are high compared to the frequency scale relevant for the overdamped dynamics of the system.
\end{abstract}

\maketitle

\section{Introduction}

In classical and statistical physics, the overdamped limit is an extremely useful
approximation that allows to simplify problems where the dynamics of a
system is dominated by the friction due to its interaction with an environment.
This can be understood based on the canonical example of a Brownian particle, where the limit of strong friction induces a time scale separation in which the
momentum degree of freedom relaxes much faster that the position. In such a case,
the Fokker-Planck equation describing the stochastic evolution of both degrees
of freedom can be reduced to the Smoluchowski equation for the evolution of
the probability density of the position alone \cite{hannes1996}. 

For quantum systems, an analogous procedure proves to be more demanding. This is
because in general tractable descriptions for the reduced dynamics of open quantum systems can only be obtained for weak coupling between the system and the environment. In contrast, by definition, the overdamped limit is a strong coupling regime (however, this does not prevent weak coupling master equations from providing approximate descriptions of overdamped dynamics under some conditions \cite{esposito2005}). In spite of this, a quantum version of the Smoluchowski equation was first obtained by Ankerhold and collaborators in \cite{pechukas2000, ankerhold2001} using path integral techniques. Those results, as well as later extensions to time-dependent systems \cite{dillenschneider2009}, only consider equilibrium environments, i.e., the system in question only interacts with a single thermal bath.

\begin{figure}[t]
  \centering
  \includegraphics[scale=.8]{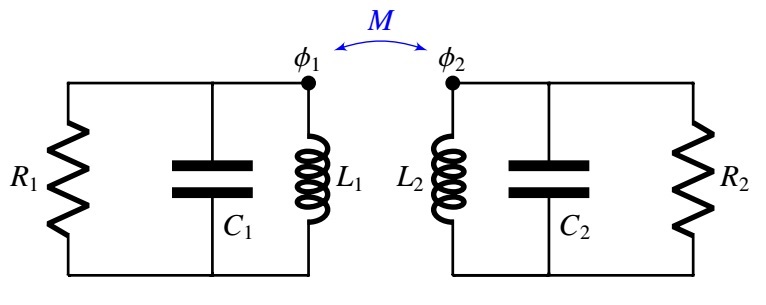}
    \caption{Two magnetically coupled RLC circuits.}
    \label{fig:model}
  \end{figure}

More recently, some efforts in stochastic and quantum thermodynamics
have also focussed on understanding the impact of strong coupling effects,
both in equilibrium and out of equilibrium settings
\cite{esposito2015,esposito10,esposito15,bergmann20,bruch16,katz2016,seifert2016,freitas2017,perarnau2018,bruch18,dou2018,haughian18,strasberg2019}.
In this article we explore the overdamped limit of a quantum system in contact with a non-equilibrium environment, i.e, we consider a situation in which the
system simultaneously interacts with two thermal baths at different temperatures.
Specifically, we consider an electrical circuit composed of two parallel RLC
circuits coupled by a mutual inductance (see Figure~\ref{fig:model}). Here,
the resistors represent the thermal baths into which energy can be dissipated.
If they are at different temperatures then the system will reach a
nonequilibrium stationary state in which heat flows from the hot to the cold
resistor. We are interested in studying the properties of this heat current
in the overdamped limit where dissipation dominates, and that in this case is achieved for $C_i R_i^2 \ll L_i$. For this purpose, we will exploit
the fact that for linear systems like this one an exact integral expression for
the heat currents can be obtained, and in some cases it can be
evaluated analytically \cite{martinez2013, freitas2014}.
In this way, we are able to split the heat current into classical and
quantum contributions and to analyse their behaviour in different regimes.
We obtain analytical expressions for both contributions that fully take
into account non-Markovian effects.
Interestingly, we show that the quantum corrections to the heat current do not necessarily
vanish in the limit where both temperatures are high with respect
to the slow frequency scale (the only one relevant for the dynamics of the
circuit in the overdamped regime). The surviving quantum corrections are non-trivial
and depend logarithmically on the temperatures. We show that these results are
indeed accurate by comparing them to exact numerical computations.

This article is organized as follows. In Sec.~\rom{2},
we describe our model of quantum circuits and introduce
the expression for the steady-state heat currents for
general harmonic networks. Next, in Sec.~\rom{3} we give the result for the classical
and quantum contributions to the steady-state heat current
in terms of the circuit parameters.
The evaluation of the heat currents in the overdamped regime is then done in Sec.~\rom{4} together with the analysis of different regimes.



\section{The model and its solution}

We begin by building a quantum model of the circuit in Fig.~\ref{fig:model}.
For this, we will represent each resistor using the Caldeira-Leggett model.
In this model, a resistor is considered as an infinite array of independent LC circuits or harmonic modes.
In this way, by the usual procedure for canonical quantization \cite{devoret2016,girvin2014},
it is possible to obtain the following Hamiltonian for the full system
(see Appendix \ref{ap:hamiltonian} for more details):
\begin{align}
H&=
\frac{q_{1}^{2}}{2C_{1}}\!+\!\frac{q_{2}^{2}}{2C_{2}}
\!+\!\frac{L_1L_2}{L_{1}L_{2}\!-\!M^{2}}
\left(\frac{\phi_{1}^{2}}{2L_{1}}\!+\!
\frac{\phi_{2}^{2}}{2L_{2}}\!-\!
\frac{M}{L_{1}L_{2}}\phi_{1}\phi_{2}\right)\nonumber\\
&+\!\sum_{m_{1}}
\left(\!\frac{q_{m_{1}}^{2}}{2C_{m_{1}}}\!+\!
\frac{(\phi_{m_{1}}\!-\!\phi_{1})^{2}}{2L_{m_{1}}}\!\right)\!+\!
\!\sum_{m_{2}}
\left(\!\frac{q_{m_{2}}^{2}}{2C_{m_{2}}}\!+\!
\frac{\!(\phi_{m_{2}}\!-\!\phi_{2})^{2}}{2L_{m_{2}}}\!\right).
	\end{align}
Here, $q_1$ and $q_2$ are quantum mechanical operators associated with the charge
on the capacitors, while $\phi_1$ and $\phi_2$ are operators associated with the
total magnetic flux through the inductors. They satisfy the usual commutation
relations $[q_j, \phi_k] = i\hbar \delta_{j,k}$. In a similar way,
$\{q_{m_1}, \phi_{m_1}\}$ and $\{q_{m_2}, \phi_{m_2}\}$ are sets of conjugate
operators associated with each of the individual modes in the Caldeira-Leggett
model of each resistor. These individual modes are characterized by
capacitances $C_{m_k}$ and inductances $L_{m_k}$, that are in principle arbitrary.
They enter in the definition of the spectral density associated to the resistors,
defined below.

In the following it will be convenient to write the different terms of the previous
Hamiltonian in matrix form. In fact, we can write
\begin{equation}
H = H_\text{sys} + \sum_{\alpha} \left( H_{\text{env}, \alpha} + H_{\text{int}, \alpha} \right),
\end{equation}
with
\begin{align}
	H_\text{sys}&=  q^{T}\frac{C^{-1}}{2}q + \phi^{T}\frac{L^{-1}}{2} \phi, \\
	H_{\text{env}, \alpha}&= q_{\alpha}^{T}\frac{C_{\alpha}^{-1}}{2}q_{\alpha} +
  \phi_{\alpha}^{T}\frac{L_{\alpha}^{-1}}{2}\phi_{\alpha}, \\
	H_{\text{int}, \alpha}&=-\phi^{T}\bar{L}_{\alpha}^{-1}\phi_{\alpha}.
\end{align}
Here, $H_\text{sys}$, $H_{\text{env},\alpha}$ and $H_{\text{int},\alpha}$
are the Hamiltonians of the system, the thermal baths and the interaction
between the system and the baths, respectively.
The index $\alpha\in\{1,2\}$ identifies a resistor or thermal bath in the environment. Moreover, $q=( q_{1},   q_{2})^T$ and
$\phi= (\phi_{1}, \phi_{2})^T$
are column vectors of the charge and flux operators of the system, respectively,
and the matrices appearing in $H_\text{sys}$ are defined by
\begin{equation}
C=\begin{pmatrix}
C_{1} & 0 \\
0 & C_{2}
\end{pmatrix} \text{,} \qquad
L_0 =\begin{pmatrix}
L_{1} & -M \\
-M & L_{2}
\end{pmatrix},
\end{equation}
and
\begin{equation}
L^{-1}=
L_0^{-1}
+\begin{pmatrix}
	\sum_{m_1}L^{-1}_{m_1} & 0 \\
	0 & \sum_{m_2}L^{-1}_{m_2}
	\end{pmatrix}.
\end{equation}
In a similar way, $q_{\alpha}$ and $\phi_{\alpha}$ are column vectors
formed with the charge and flux operators of the $\alpha$-th bath,
and $C_{\alpha}$ and $L_{\alpha}$ are diagonal matrices containing the capacitances
and inductances of each bath. Finally, the matrices $\bar{L}^{-1}_{\alpha}$
are given by
\begin{align}
 \bar{L}^{-1}_{1}=
	\begin{pmatrix}
	L_{1 1} & L_{1 2} && ... & L_{1N}\\
	0 & 0 && ...&0
	\end{pmatrix}, \\
	\bar{L}^{-1}_{2}=
	\begin{pmatrix}
	0 & 0 && ...&0\\
	L_{2 1} & L_{2 2} && ... & L_{2N}
	\end{pmatrix}.
\end{align}

The system described so far is a particular case of a open harmonic network.
The non-equilibrium thermodynamics of these systems has been extensively studied
before \cite{esposito10, esposito15, gaul2007, asadian2013, martinez2013, freitas2014, nicacio2015, freitas2015, freitas2017}, since owing to their
linearity exact analytical results can be obtained. The central quantities in this
study will be the heat currents associated to each thermal bath, i.e., the
average rates at which energy is interchanged between the system and each bath.
They can be defined as (heat currents are considered positive when they enter the system):
\begin{equation}
\dot Q_{\alpha} = - \frac{1}{i \hbar} \meann{[H_{\text{env},\alpha}, H_{\text{int},\alpha}]},
\end{equation}
where the mean value is taken with respect to the instantaneous global state.
Given an initial state, the heat currents $\dot Q_{\alpha}$ will
depend nontrivially on time during relaxation, after which they will reach
stationary values. Typically, one assumes an
uncorrelated initial state $\rho_0=\rho_\text{sys}\otimes\rho_\text{env}$
in which each of the baths in the environment is in a
thermal state $\rho_\alpha^\text{th}$ at inverse temperature
$\beta_\alpha = (k_b T_\alpha)^{-1}$, i.e.,
$\rho_\text{env} = \otimes_\alpha \rho_\alpha^\text{th}$.
Under this assumption, it can be shown that in the long-time limit the average heat currents can be expressed as (see App.~\ref{ap:heat_current_proof}),
\begin{equation}
\begin{split}
\dot{Q}_{\alpha} =  \frac{\hbar}{2} \sum_{\alpha^{\prime}\neq\alpha}
& \int_{0}^{\infty}
\!\!  d\omega \, \omega \, f_{\alpha\alpha^{\prime}}(\omega) \\
& \times
\left(
\coth\left(\beta_{\alpha}\hbar\omega/2\right)
-
\coth\left(\beta_{\alpha'}\hbar\omega/2\right)
\right),
\label{eq:heat_current}
\end{split}
\end{equation}
where $f_{\alpha\alpha^{\prime}}(\omega)$ is the heat
transfer matrix element and reads
\begin{equation}\label{eq:heat_matrix}
f_{\alpha\alpha^{\prime}} (\omega)=\frac{\pi}{2}\mathrm{Tr}
\left[I_{\alpha}(\omega)g(i\omega)I_{\alpha^{\prime}}(\omega)
g^{\dagger}(i\omega)\right].
\end{equation}
In the previous expression, $I_\alpha(\omega)$ is the spectral density of
the $\alpha$th bath. It is a $2\times 2$ matrix with elements
\begin{equation}
[I_{\alpha}(\omega)]_{kl} = \sum_{n} \! \left(\bar{L}_{\alpha}^{-1}\right)_{kn}  (\bar{L}_{\alpha}^{-1})_{ln}(\omega_{\alpha} C_{\alpha})_{nn}^{-1}\delta\left[\omega- \left(\omega_{\alpha}\right)_{nn}\right], \end{equation}
where $\omega_{\alpha}^2=L^{-1}_{\alpha}C^{-1}_{\alpha}$ is a diagonal matrix with
the squared natural frequencies of the modes in the $\alpha$th bath.
Also, $g(s)$ in Eq.~\eqref{eq:heat_matrix} is the Laplace transform of the circuit
Green's function,
\begin{equation}
g(s)^{-1}=Cs^2+\gamma(s)s+L^{-1}_{0},
\label{eq:green}
\end{equation}
where $\gamma(s)$ is the Laplace transform of the dissipation kernel. It is
given by
\begin{equation}
\gamma(s)=\int_{0}^{\infty}\frac{I(\omega)}{\omega} \frac{s}{s^2 + \omega^2} d\omega
\label{eq:gamma}
\end{equation}
in terms of the total spectral density $I(\omega) = \sum_\alpha I_\alpha(\omega)$.

The frequency integral in Eq.~\eqref{eq:heat_current} can be solved analytically
in certain cases. As shown in Ref.~[\onlinecite{freitas2014}], when the spectral densities of
all baths are of the Lorentz-Drude form, the integral can be evaluated in terms
of the eigenvalues and eigenvectors of a cubic eigenvalue problem. Thus,
we will assume the following spectral density for the baths:
\begin{equation}
I_{\alpha}(\omega)=\frac{2}{\pi}\frac{1}{R_\alpha}
\frac{\omega\:\omega_{c}^2}{\omega^2+\omega_{c}^2}P_{\alpha},
\end{equation}
with
\begin{equation}
	P_1=\begin{pmatrix}
	1 & 0\\
	0 & 0
  \end{pmatrix}\:\:\: \text{   and   } \:\:\:
	P_2=\begin{pmatrix}
	0 & 0\\
	0 & 1
	\end{pmatrix}.
\end{equation}
Interestingly, the previous results are valid for any value of the cutoff frequency
$\omega_c$, which controls the autocorrelation time of the environment (the Markovian approximation corresponds to the limit $\omega_c \to \infty$). Thus,
our results will automatically include non-Markovian effects.
Finally, with the previous choice for the spectral densities, the function $\gamma(s)$
in Eq.~\eqref{eq:gamma} becomes
\begin{equation}
\gamma(s) = \left( \frac{P_1}{R_1} + \frac{P_2}{R_2} \right) \frac{\omega_c}{s + \omega_c}.
\end{equation}

\section{Classical and quantum contributions to the heat current}

The previous ingredients enable us to find the heat current in terms of
the circuit parameters. By plugging the definitions of the spectral densities
into Eq.~\eqref{eq:heat_matrix} we find
\begin{align}
f_{1,2}(\omega) = \frac{2}{\pi}
\left(\frac{1}{R}
\frac{\omega\:\omega_{c}^2}{\omega^2+\omega_{c}^2}\right)^{2}
 |g_{12}(i\omega)|^{2},
\end{align}
where $g_{1,2}(s)$ is the off-diagonal element of $g(s)$.
For simplicity we will consider the case of a symmetric circuit, i.e.,
$R_1 = R_2 = R$, $C_1=C_2=C$, and $L_1=L_2=L$. Then, we obtain from Eq.~\eqref{eq:green},
\begin{align}
g_{12}(s)=\frac{M}{A}
\left[\left(C s^{2}+\frac{L}{A}+
\frac{1}{R}\frac{ s \: \omega_{c}}{ s+\omega_{c}}\right)^{2}
-\left(\frac{M}{A}\right)^{2}\right]^{-1},
\end{align}
where $A = L^2 - M^2$. As a consequence, the transfer function can be finally written as
\begin{equation}
f_{1,2}(\omega)=\frac{2}{\pi}\: \omega^2\omega_{c}^4
\left(\frac{RM}{A}\right)^2\frac{1}{|u_{+}(i\omega)\: u_{-}(i\omega)|^2},
\end{equation}
with
\begin{equation}
u_{\pm}(s)=(s^3+\omega_{c}s^2)RC \:\:+\:\:
\left(\frac{R}{L\pm M}+ \omega_{c}\right)s \:\:+\:\:
\frac{R}{L\pm M}\omega_{c}.
\label{eq:def_u}
\end{equation}
We can already
see how an exact expression for the heat current can be obtained.
Since the transfer function $f_{1,2}(\omega)$ was expressed
as a rational function, the frequency integral in Eq.~\eqref{eq:heat_current}
can be evaluated via the residue theorem in terms of the poles
of $f_{1,2}(\omega)$. In order to deal with the poles of the
functions $\coth(\beta_\alpha \hbar \omega/2)$ at the Matsubara frequencies, it is convenient to write them in terms of digamma functions \cite{riseborough1985} (see App.~\ref{ap:quantum_corrections}):
\begin{align}
	\pi \coth\left(\frac{\beta_{\alpha}\hbar \omega}{2}\right)=
  \frac{2\pi}{\beta_{\alpha}\hbar\omega}&-i\psi\left(1-\frac{i\beta_{\alpha}\hbar \omega}{2\pi}\right)\notag \\ &+i\psi\left(1+\frac{i\beta_{\alpha}\hbar \omega}{2\pi}\right).
\end{align}
We note that the terms containing digamma functions vanish in the high-temperature
limit. Thus, this decomposition induces a splitting of the heat current
into a high-temperature contribution and a low-temperature correction, which
we denote as classical and quantum contributions, respectively. Therefore, we have
\begin{equation}
\dot Q_1 = \dot Q_1^\text{cl} + \dot Q_1^\text{q},
\end{equation}
where
\begin{align}
\dot{Q}_{1}^\text{cl}&=\left(\frac{1}{\beta_{1}}-
\frac{1}{\beta_{2}}\right)\int_{0}^{\infty}d\omega f_{12} (\omega) , \label{eq:classical_heat} \\
\dot{Q}_{1}^\text{q}&=
\frac{i\hbar}{2}  \int_{-\infty}^{\infty}  d\omega \omega
f_{12}(\omega)\left[\psi\left(1-\frac{i\beta_{2}\hbar \omega}{2\pi}\right)
-\psi\left(1-\frac{i\beta_{1}\hbar \omega}{2\pi}\right)\right].
\label{eq:quantum_heat}
\end{align}
Although the previous integrals could in principle be evaluated exactly \cite{freitas2014}, the procedure and the final result are greatly simplified
in the overdamped limit in which we are interested. Thus, we now discuss
this approximation and the frequency scales involved.

\section{Evaluation of the heat current in the overdamped limit}

The classical equation of motion for a single parallel RLC circuit is
\begin{equation}
\ddot{\phi}+\gamma\dot{\phi}+\omega_{0}^2\phi=0,
\end{equation}
where $\phi$ is the flux variable in the inductor, and the relevant frequency
scales are given by the damping rate $\gamma=1/RC$ and the natural frequency
$\omega_0=1/\sqrt{LC}$. The corresponding characteristic equation has roots
$\Gamma_{\pm}=-(\gamma/2)\pm (\gamma^2/4-\omega_0^2)^{1/2}$.
The overdamped limit corresponds to $\gamma \gg \omega_0$, and it can be
reached for instance by reducing the value of the capacitance
so that ${C \ll L/R^2}$. In that regime we have
$\Gamma_+\simeq-\omega_0^2/\gamma$ and $\Gamma_-\simeq-\gamma+\omega_0^2/\gamma$,
that in absolute value are
the damping rates of the magnetic flux $\phi$ and charge $q$, respectively,
and therefore $|\Gamma_+| \ll |\Gamma_-|$.
This is the time-scale separation characteristic for overdamped systems, which means in this case that the charge relaxes much faster than the flux. Also, note that the flux damping rate
$\omega_d \simeq \omega_0^2/\gamma =  R/L$ becomes independent of $C$.
A similar analysis holds for each normal mode of the two coupled RLC circuits
by just replacing $L$ by $L\pm M$.
We can express the functions $u_\pm(s)$ in Eq.~\eqref{eq:def_u} in terms
of $\gamma$ and $\omega_\pm = \omega_d/(1 \pm M/L)$,
\begin{equation}
u_\pm(s)= (s^3+ \omega_c s^2)/\gamma+
\left(\omega_\pm+\omega_{c}\right)s+\omega_\pm\omega_{c}.
\label{eq:def_u_timescales}
\end{equation}

We see that the overdamped limit tends to reduce the weight of the cubic and
quadratic terms, although they will always dominate for high frequencies.
However, we also note that in the frequency integral of Eq.~\eqref{eq:heat_current}, the factor
$
\coth\left(\beta_{\alpha}\hbar\omega/2\right)
-
\coth\left(\beta_{\alpha'}\hbar\omega/2\right)
$ will cut off frequencies higher than
$\omega_\text{th} =k_b\max_\alpha\{T_\alpha\}/\hbar$.
From this it follows that the cubic and quadratic terms can be disregarded
with respect to the other two whenever
\begin{table}
\begin{tabular}{|| l  c r ||}
\hline
(a) High temperatures: & $\qquad$ &  $\omega_\pm \ll \gamma \ll \omega_\text{th}$\\
(b) Intermediate temperatures: & $\qquad$ & $\omega_\pm < \omega_\text{th} \ll \gamma$\\
(c) Low temperatures: & $\qquad$ & $\omega_\text{th} < \omega_\pm \ll \gamma$\\
\hline
\end{tabular}
\caption{Different temperature ranges in the overdamped regime. We consider that
the thermal frequency $\omega_\text{th}$ characterizes the temperatures of both baths,
i.e., both the temperatures are of the same order. (a) is the range adressed by the classical
Smoluchowski equation or overdamped Langevin equations.
In (b), the bath temperatures sit in between
the frequency gap associated to the overdamped regime, while in (c) temperatures
are low compared to the lowest frequency scale of the system.
Other ranges can be considered, for example mixed conditions in which one
of the bath temperatures is low while the other is high, or taking into account
values of $\omega_\text{th}$ comparable to $\gamma$.}
\label{tb:regimes}
\end{table}

\begin{equation}
\omega_\text{th} \ll \gamma, (\gamma \omega_\pm)^{1/2}, (\gamma \omega_\pm
\omega_c)^{1/3}.
\label{eq:over_conditions}
\end{equation}
Thus, under those conditions, we can consider
	\begin{equation}
	u_\pm(s) \simeq
  \left(\omega_\pm+\omega_{c}\right)s+\omega_\pm\omega_{c},
  \label{eq:def_u_approx}
	\end{equation}
where the only remaining relevant frequency scales are $\omega_\pm$ and $\omega_c$.
We note that the conditions in
Eq.~\eqref{eq:over_conditions} can always be fulfilled
by increasing $\gamma$, and that they do not restrict in any way the ratios
between $\omega_\text{th}$, $\omega_\pm$, and $\omega_c$. However, they pose a restriction on the maximum value of the temperatures and Table~\ref{tb:regimes} specifies some temperature ranges relevant in the overdamped regime.
This will become important later when we show that quantum effects survive
even when the temperatures are high with respect to $\hbar \omega_\pm/k_b$.
\begin{figure}[!ht]
	\includegraphics[width=.48\textwidth]{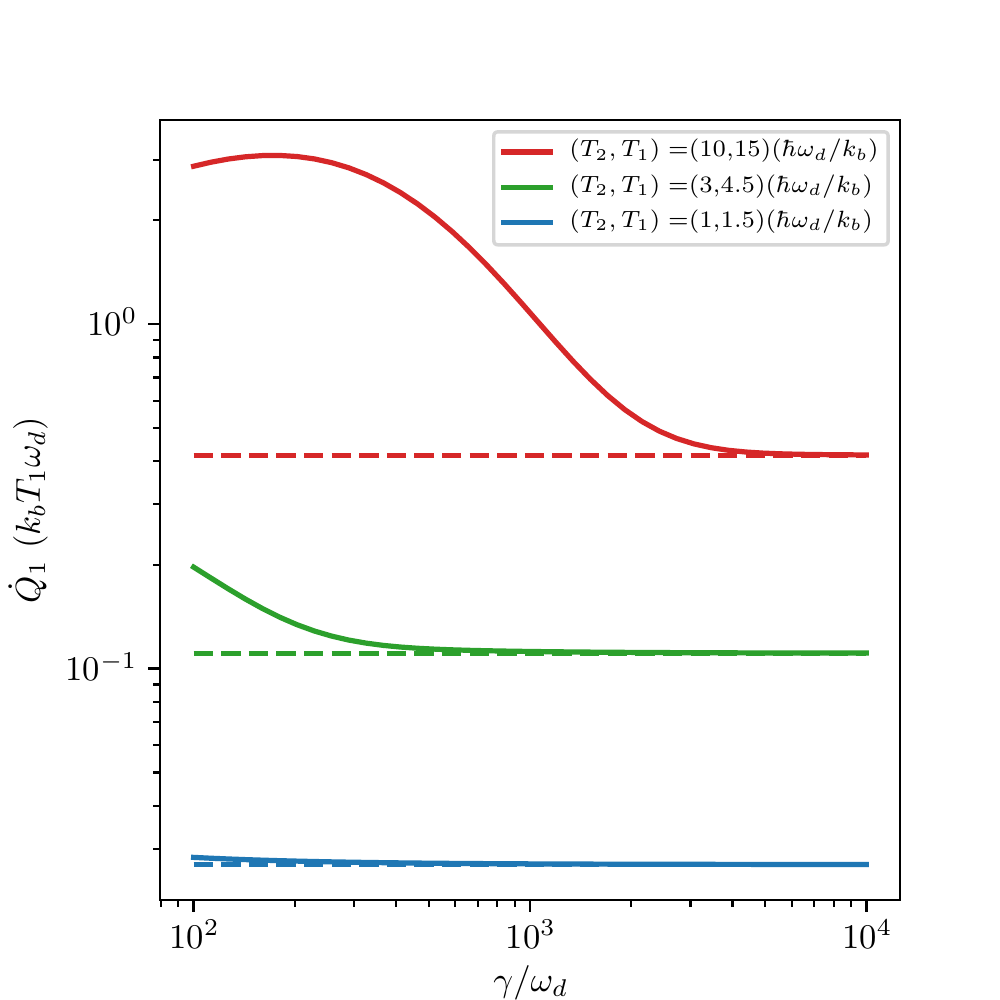}
	\caption{Heat current with respect to $\gamma/\omega_d$.
  Solid lines show the exact heat current in terms of the different values
  of baths temperatures $T_1$ and $T_2$. The dashed lines are the heat
  current in the overdamped limit. This plot is sketched for $M=1,L=2,\omega_c=5\omega_d,\omega_d=1$. }
      \label{fig:exact_vs_overdamped}
\end{figure}
Using the approximation of Eq.~\eqref{eq:def_u_approx},
the integrals in Eqs.~\eqref{eq:classical_heat} and \eqref{eq:quantum_heat}
can be readily evaluated. For the classical contribution to the heat current,
we obtain:

\begin{equation}
\dot{Q}^\text{cl}_{1}=\frac{k_b}{2}
\left(T_1-T_2\right)
\left(\frac{M}{L}\right)^2
\frac{\omega_c}{\omega_c+\omega_d} \frac{\lambda_+ \lambda_-}{\omega_d},
\label{eq:classical_contr}
\end{equation}
where $\lambda_\pm$ is the only root of $u_\pm(s)$,
\begin{equation}
\lambda_\pm = -\frac{\omega_c\: \omega_\pm}{\omega_c+\omega_\pm}.
\label{eq:roots}
\end{equation}
The evaluation of the quantum contribution is not as straightforward,
and its details are explained in App.~\ref{ap:quantum_corrections}.
The final result is


\begin{widetext}
\begin{align}
\dot{Q}^\text{q}_{1}&=
\frac{\hbar}{\pi}
\left(\frac{M}{L}\right)^2
\left(\frac{\lambda_+\lambda_-}{\omega_d}\right)^2
\log\left(\frac{T_2}{T_1}\right)\nonumber\\
&+\frac{\hbar}{4\pi}
\frac{\omega_c}{\omega_c+\omega_d}
\frac{M}{L}
\Bigg\{
\lambda_+^2
\left[
\psi\left(1-\frac{\beta_1\hbar\lambda_+}{2\pi}\right)-
\psi\left(1-\frac{\beta_2\hbar\lambda_+}{2\pi}\right)
\right]
-\lambda_-^2
\left[
\psi\left(1-\frac{\beta_1\hbar\lambda_-}{2\pi}\right)-
\psi\left(1-\frac{\beta_2\hbar\lambda_-}{2\pi}\right)
\right]
\Bigg\}.
\label{eq:quantum_contr}
\end{align}
\end{widetext}
Equations~\eqref{eq:classical_contr} and \eqref{eq:quantum_contr} are the central
results of this work. They make it possible to compute the heat current in the overdamped
regime without assuming the weak coupling or Markovian approximations, and thus
complement previous results in similar systems that are either numerical or
limited by the mentioned approximations \cite{gaul2007, asadian2013, nicacio2015, freitas2015}.
\begin{figure}[!hb]
\centering
     \includegraphics[width=.48\textwidth]{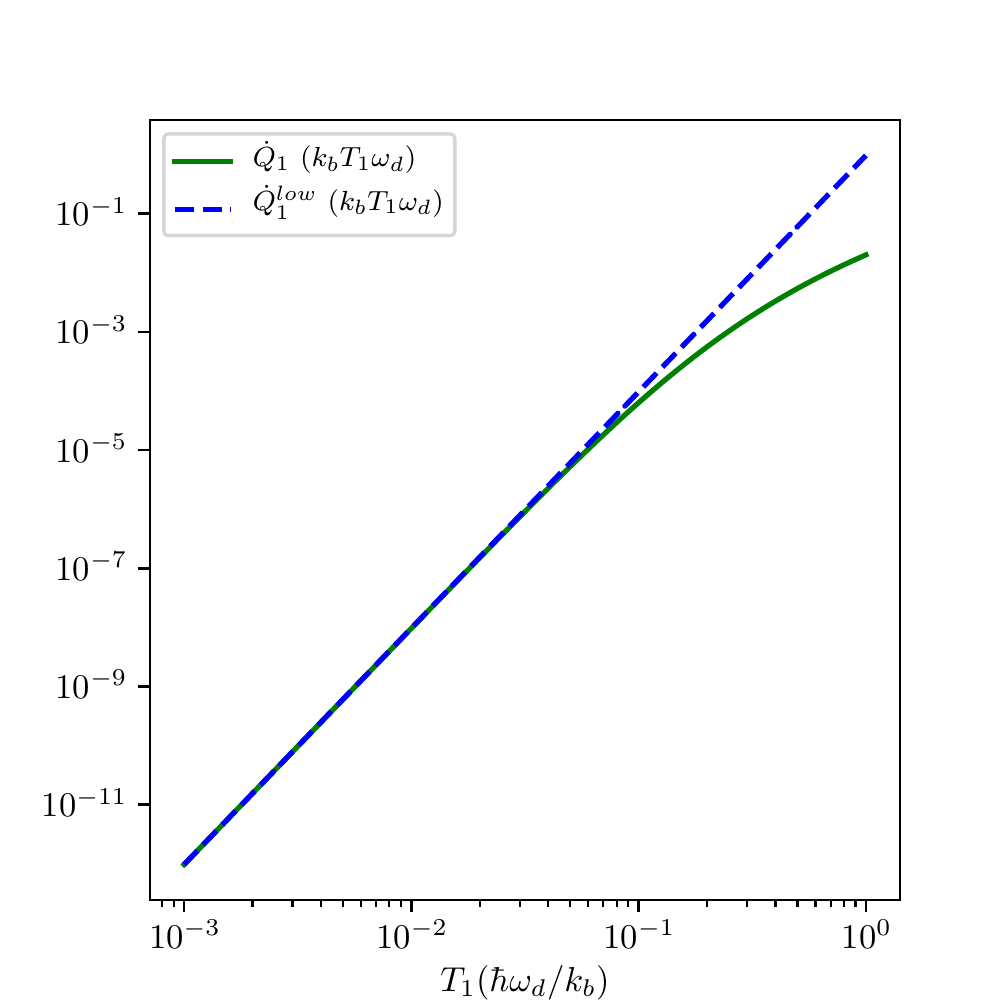}
	\caption{Comparison between total heat current (green solid line) and the low temperature regime expression (blue dashed line) with respect to the different values of $T_1$. Here we set $T_2=T_1/2 $ thus we lower the two temperatures at the same time with a constant ratio. We can observe that, when the temperatures are lowered, the two expressions will coincide. ($M=1,L=2,\omega_c=5\omega_d,\omega_d=1$).}
      \label{fig:low_temp}
\end{figure}
We observe that the classical contribution is proportional to the temperature
difference $\Delta T = T_1 - T_2$, whereas the quantum contribution
depends on $T_1$ and $T_2$ in a non-algebraic way, as expected.
In Fig.~\ref{fig:exact_vs_overdamped} we compare the exact heat current obtained
by numerical integration of Eq.~\eqref{eq:heat_current} with the one obtained
by using Eqs.~\eqref{eq:classical_contr} and \eqref{eq:quantum_contr},
for different values of
$T_1$ and $T_2$, as a function of $\gamma/\omega_d$ ($M/L=1/2$ and $\omega_c = 5 \omega_d$).
We see that the two results match as $\gamma/\omega_d$ is increased.


We will now take some relevant limits in order to simplify the previous
expressions.
The Markovian limit ($\omega_c \to \infty$) can be easily obtained by replacing
the factors $\omega_c/(\omega_c+\omega_d)$ by $1$ in Eqs.~\eqref{eq:classical_contr} and \eqref{eq:quantum_contr} and noting that the
roots $\lambda_\pm$ satisfy
\begin{equation}
\lim_{\omega_c \to \infty} \lambda_\pm = -\omega_\pm.
\end{equation}
From Eq.~\eqref{eq:roots} we see that the effect of a finite cutoff
is equivalent to reducing the values of the frequencies $\omega_\pm$ or, correspondingly, $\omega_d$.

To analyze the low-temperature regime we consider the limit $|\lambda_\pm|/\omega_\text{th} \gg 1$
(note that this condition implies ${\omega_\text{th} \ll \omega_\pm, \omega_c}$). We
use the following expansion of the digamma function for large $x$,
\begin{equation}
	\psi(x)\approx
  \log x-\frac{1}{2x}-\frac{1}{12x^2}+\frac{1}{120x^4}+\mathcal{O}(x^5),
  \label{eq:digamma_low_temp}
\end{equation}
The contribution of the first logarithmic term cancels the first term in Eq.~\eqref{eq:quantum_contr}. Moreover, the contribution
of the second term cancels the classical part of the heat current,
while those coming from the third term vanish. Thus, the only
remaining contributions originate from the term $\propto 1/x^4$, so the
final result for the low-temperature heat current is
\begin{figure}[!t]
\centering
	\includegraphics[width=.48\textwidth]{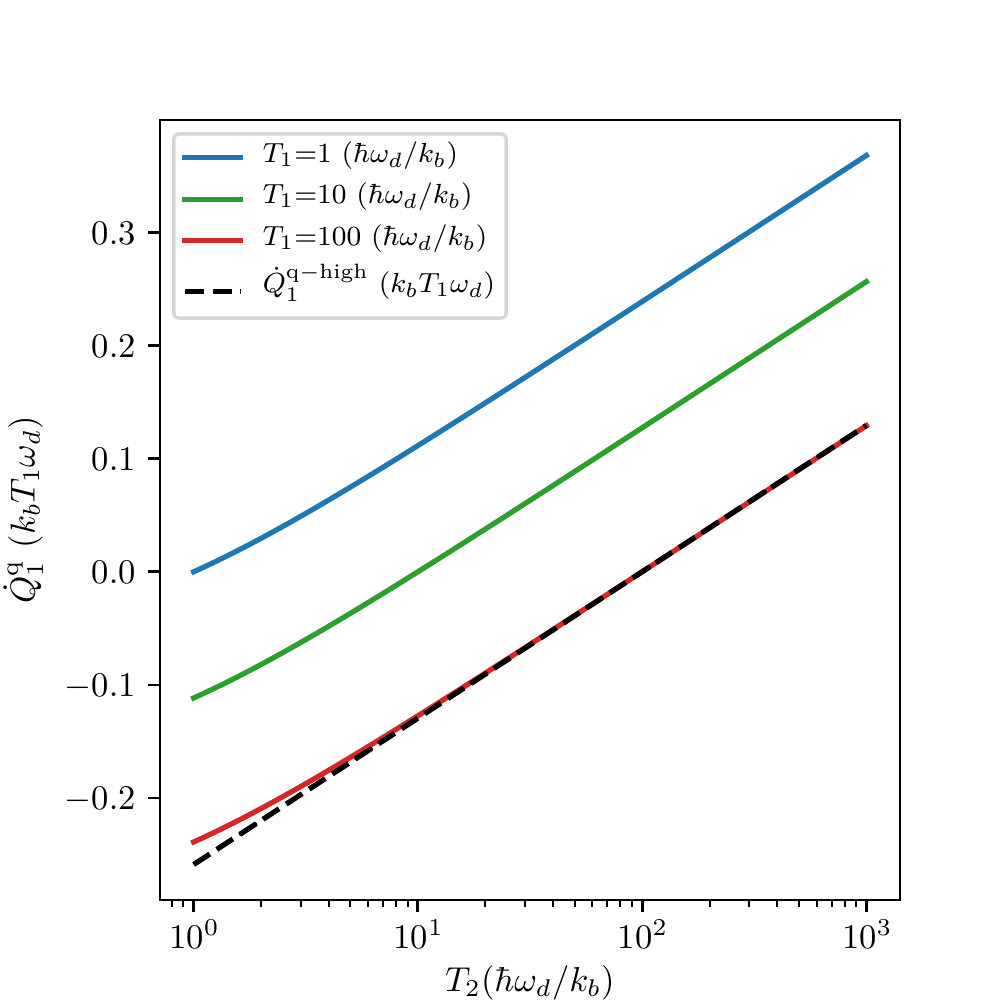}
			\caption{The quantum contribution to the heat currents for different values of $T_1$ is sketched with respect to $T_2$. We can see that, for high values of $T_1$ and $T_2$, the quantum correction is not vanishing and it will coincide with a non-trivial logarithmic expression (black dashed line). This plot is sketched for $M=1,L=2,\omega_c=5\omega_d,\omega_d=1$. }
      \label{fig:high_temp}
      \end{figure}
\begin{equation}
\dot{Q}_{1}^\text{low}=\frac{2}{15}\left(\frac{\pi}{\hbar}\right)^3
\left(\frac{M}{L}\right)^2\frac{k_b^4}{\omega_d^2}\left(T_1^4-T_2^4\right)
+\mathcal{O}(T_{1/2}^6)
.
\label{eq:heat_low_temp}
\end{equation}
Due to the scaling with temperature, this expression is reminiscent of the Stefan-Boltzmann law for black-body
radiation, which is a thermal equilibrium result. Indeed, a similar result would be obtained
by considering two black bodies at thermal equilibrium but different temperatures which radiate towards each other. Thus, we see that
non-equilibrium effects appear only at next-to-leading order in the low-temperature regime and are fully captured by Eq.~\eqref{eq:quantum_contr}.
Also, we would like to point out that Eq.~(\ref{eq:heat_low_temp}) is independent of the cutoff frequency. This is natural since for low
temperatures only low frequency modes contribute to the heat current, while
the cutoff frequency controls the high-frequency region of the spectral densities.
In Fig.~\ref{fig:low_temp} we sketch the behavior of the total heat current when the bath temperatures are decreased. For low temperatures, we can see that the heat current is indeed well approximated by Eq.~\eqref{eq:heat_low_temp}.

Turning to the regime of intermediate temperatures, where
$|\lambda_\pm|/\omega_\text{th} \ll 1$, we employ the following expansion
of the digamma function for small values of $x$,

\begin{equation}
\psi\left(1+x\right)=-\eta+\frac{\pi^2x}{6}+\mathcal{O}(x^2),
\end{equation}
where $\eta$ is the Euler–Mascheroni constant. We then find the following
high-temperature expansion of the quantum contribution

\begin{equation}
\begin{split}
\dot{Q}^\text{q}_{1}&=
\frac{\hbar}{\pi}
\left(\frac{M}{L}\right)^2
\left(\frac{\lambda_+\lambda_-}{\omega_d}\right)^2
\log\left(\frac{T_2}{T_1}\right)\\
&+\frac{\hbar^2}{48}
\frac{\omega_c}{\omega_c+\omega_d}
\frac{M}{L} (\lambda_+^3 - \lambda_-^3) \left(\frac{1}{T_2} - \frac{1}{T_1}\right)
+\mathcal{O}(T_{1/2}^{-2}).
\label{eq:heat_quantum_high}
\end{split}
\end{equation}
Surprisingly, we see that the dominant term does not necessarily
vanish for $|\lambda_\pm|/\omega_\text{th} \to 0$. The reason for this is that under the constraints given in Eq.~\eqref{eq:over_conditions}, one can assume that the temperature is high compared to the slow frequency
scale $\omega_d$, but it must remain low compared to the fast frequency scale $\gamma$.
In other words, the temperature sits in the middle of the
time scale separation associated to the overdamped regime.
Thus, to the first non-trivial order the total heat current for high temperatures is,
\begin{align}
\dot Q^\text{high}_1 = \dot Q^\text{cl}_1 +
\frac{\hbar}{\pi}
\left(\frac{M}{L}\right)^2
\left(\frac{\lambda_+\lambda_-}{\omega_d}\right)^2
\log\left(\frac{T_2}{T_1}\right).
\label{eq:heat_high_temp}
\end{align}
Figure~\ref{fig:high_temp} shows the behavior of the quantum contribution with respect to the growth of the temperature. When both bath temperatures are increased, we can still observe a non-zero quantum correction to the heat currents.

\section{Conclusions}

We have investigated the heat current between two overdamped quantum
harmonic oscillators interacting with local thermal baths, without invoking the
weak coupling and Markovian approximations. Exploiting the time-scale separation
associated to the overdamped regime we were able to obtain closed analytical expressions
for the heat current, identifying quantum and classical contributions. These analytical
results might offer a useful benchmark to test Markovian embedding schemes or other approximate
methods, for example the one developed in \cite{Mascherpa2020}.
Although our results are valid for
general harmonic systems, we have explicitly considered an electronic implementation.
This is justified by the fact that low-temperature electronic circuits are a promising platform
to study quantum energy transport \cite{Pascal2011,Pekola2015,Ronzani2018,Li2019}.
We found that in the overdamped regime a range of
intermediate temperatures opens up between the low-temperature
and high-temperature regimes usually considered. Our results indicate that in this intermediate range
there are significant quantum corrections to the classical heat current,
which survive even if the temperatures are high compared to the only relevant
frequency scale of the system dynamics.
\bibliography{references}

\begin{thebibliography}{32}%
\makeatletter
\providecommand \@ifxundefined [1]{%
 \@ifx{#1\undefined}
}%
\providecommand \@ifnum [1]{%
 \ifnum #1\expandafter \@firstoftwo
 \else \expandafter \@secondoftwo
 \fi
}%
\providecommand \@ifx [1]{%
 \ifx #1\expandafter \@firstoftwo
 \else \expandafter \@secondoftwo
 \fi
}%
\providecommand \natexlab [1]{#1}%
\providecommand \enquote  [1]{``#1''}%
\providecommand \bibnamefont  [1]{#1}%
\providecommand \bibfnamefont [1]{#1}%
\providecommand \citenamefont [1]{#1}%
\providecommand \href@noop [0]{\@secondoftwo}%
\providecommand \href [0]{\begingroup \@sanitize@url \@href}%
\providecommand \@href[1]{\@@startlink{#1}\@@href}%
\providecommand \@@href[1]{\endgroup#1\@@endlink}%
\providecommand \@sanitize@url [0]{\catcode `\\12\catcode `\$12\catcode
  `\&12\catcode `\#12\catcode `\^12\catcode `\_12\catcode `\%12\relax}%
\providecommand \@@startlink[1]{}%
\providecommand \@@endlink[0]{}%
\providecommand \url  [0]{\begingroup\@sanitize@url \@url }%
\providecommand \@url [1]{\endgroup\@href {#1}{\urlprefix }}%
\providecommand \urlprefix  [0]{URL }%
\providecommand \Eprint [0]{\href }%
\providecommand \doibase [0]{http://dx.doi.org/}%
\providecommand \selectlanguage [0]{\@gobble}%
\providecommand \bibinfo  [0]{\@secondoftwo}%
\providecommand \bibfield  [0]{\@secondoftwo}%
\providecommand \translation [1]{[#1]}%
\providecommand \BibitemOpen [0]{}%
\providecommand \bibitemStop [0]{}%
\providecommand \bibitemNoStop [0]{.\EOS\space}%
\providecommand \EOS [0]{\spacefactor3000\relax}%
\providecommand \BibitemShut  [1]{\csname bibitem#1\endcsname}%
\let\auto@bib@innerbib\@empty
\bibitem [{\citenamefont {Hannes}\ and\ \citenamefont
  {Till}(1996)}]{hannes1996}%
  \BibitemOpen
  \bibfield  {author} {\bibinfo {author} {\bibfnamefont {R.}~\bibnamefont
  {Hannes}}\ and\ \bibinfo {author} {\bibfnamefont {F.}~\bibnamefont {Till}},\
  }\href {\doibase 10.1007/978-3-642-61544-3} {\bibfield  {journal} {\bibinfo
  {journal} {Springer-Verlag}\ } (\bibinfo {year} {1996}),\
  10.1007/978-3-642-61544-3}\BibitemShut {NoStop}%
\bibitem [{\citenamefont {Esposito}\ and\ \citenamefont
  {Haake}(2005)}]{esposito2005}%
  \BibitemOpen
  \bibfield  {author} {\bibinfo {author} {\bibfnamefont {M.}~\bibnamefont
  {Esposito}}\ and\ \bibinfo {author} {\bibfnamefont {F.}~\bibnamefont
  {Haake}},\ }\href {\doibase 10.1103/PhysRevA.72.063808} {\bibfield  {journal}
  {\bibinfo  {journal} {Phy. Rev. A}\ }\textbf {\bibinfo {volume} {72}},\
  \bibinfo {pages} {063808} (\bibinfo {year} {2005})}\BibitemShut {NoStop}%
\bibitem [{\citenamefont {Pechukas}\ \emph {et~al.}(2000)\citenamefont
  {Pechukas}, \citenamefont {Ankerhold},\ and\ \citenamefont
  {Grabert}}]{pechukas2000}%
  \BibitemOpen
  \bibfield  {author} {\bibinfo {author} {\bibfnamefont {P.}~\bibnamefont
  {Pechukas}}, \bibinfo {author} {\bibfnamefont {J.}~\bibnamefont {Ankerhold}},
  \ and\ \bibinfo {author} {\bibfnamefont {H.}~\bibnamefont {Grabert}},\ }\href
  {\doibase 10.1002/1521-3889(200010)9:9/10<794::AID-ANDP794>3.0.CO;2-J}
  {\bibfield  {journal} {\bibinfo  {journal} {Annalen der Physik}\ }\textbf
  {\bibinfo {volume} {9}},\ \bibinfo {pages} {794} (\bibinfo {year}
  {2000})}\BibitemShut {NoStop}%
\bibitem [{\citenamefont {Ankerhold}\ \emph {et~al.}(2001)\citenamefont
  {Ankerhold}, \citenamefont {Pechukas},\ and\ \citenamefont
  {Grabert}}]{ankerhold2001}%
  \BibitemOpen
  \bibfield  {author} {\bibinfo {author} {\bibfnamefont {J.}~\bibnamefont
  {Ankerhold}}, \bibinfo {author} {\bibfnamefont {P.}~\bibnamefont {Pechukas}},
  \ and\ \bibinfo {author} {\bibfnamefont {H.}~\bibnamefont {Grabert}},\ }\href
  {\doibase 10.1103/PhysRevLett.87.086802} {\bibfield  {journal} {\bibinfo
  {journal} {Phys. Rev. Lett.}\ }\textbf {\bibinfo {volume} {87}},\ \bibinfo
  {pages} {086802} (\bibinfo {year} {2001})}\BibitemShut {NoStop}%
\bibitem [{\citenamefont {Dillenschneider}\ and\ \citenamefont
  {Lutz}(2009)}]{dillenschneider2009}%
  \BibitemOpen
  \bibfield  {author} {\bibinfo {author} {\bibfnamefont {R.}~\bibnamefont
  {Dillenschneider}}\ and\ \bibinfo {author} {\bibfnamefont {E.}~\bibnamefont
  {Lutz}},\ }\href {\doibase 10.1103/PhysRevE.80.042101} {\bibfield  {journal}
  {\bibinfo  {journal} {Phys. Rev. E}\ }\textbf {\bibinfo {volume} {80}},\
  \bibinfo {pages} {042101} (\bibinfo {year} {2009})}\BibitemShut {NoStop}%
\bibitem [{\citenamefont {Esposito}\ \emph {et~al.}(2015)\citenamefont
  {Esposito}, \citenamefont {Ochoa},\ and\ \citenamefont
  {Galperin}}]{esposito2015}%
  \BibitemOpen
  \bibfield  {author} {\bibinfo {author} {\bibfnamefont {M.}~\bibnamefont
  {Esposito}}, \bibinfo {author} {\bibfnamefont {M.~A.}\ \bibnamefont {Ochoa}},
  \ and\ \bibinfo {author} {\bibfnamefont {M.}~\bibnamefont {Galperin}},\
  }\href {\doibase 10.1103/PhysRevLett.114.080602} {\bibfield  {journal}
  {\bibinfo  {journal} {Phys. Rev. Lett.}\ }\textbf {\bibinfo {volume} {114}},\
  \bibinfo {pages} {080602} (\bibinfo {year} {2015})}\BibitemShut {NoStop}%
\bibitem [{\citenamefont {Esposito}\ \emph {et~al.}(2010)\citenamefont
  {Esposito}, \citenamefont {Lindenberg},\ and\ \citenamefont
  {Broeck}}]{esposito10}%
  \BibitemOpen
  \bibfield  {author} {\bibinfo {author} {\bibfnamefont {M.}~\bibnamefont
  {Esposito}}, \bibinfo {author} {\bibfnamefont {K.}~\bibnamefont
  {Lindenberg}}, \ and\ \bibinfo {author} {\bibfnamefont {C.~V.~d.}\
  \bibnamefont {Broeck}},\ }\href {\doibase 10.1088/1367-2630/12/1/013013}
  {\bibfield  {journal} {\bibinfo  {journal} {New Journal of Physics}\ }\textbf
  {\bibinfo {volume} {12}},\ \bibinfo {pages} {013013} (\bibinfo {year}
  {2010})}\BibitemShut {NoStop}%
\bibitem [{\citenamefont {Esposito}\ and\ \citenamefont
  {Ochoa}(2015)}]{esposito15}%
  \BibitemOpen
  \bibfield  {author} {\bibinfo {author} {\bibfnamefont {M.}~\bibnamefont
  {Esposito}}\ and\ \bibinfo {author} {\bibfnamefont {M.}~\bibnamefont {Ochoa},
  \bibfnamefont {Maicol A.~andGalperin}},\ }\href {\doibase
  10.1103/PhysRevB.92.235440} {\bibfield  {journal} {\bibinfo  {journal} {Phys.
  Rev. B}\ }\textbf {\bibinfo {volume} {92}},\ \bibinfo {pages} {235440}
  (\bibinfo {year} {2015})}\BibitemShut {NoStop}%
\bibitem [{\citenamefont {Bergmann}\ and\ \citenamefont
  {Galperin}()}]{bergmann20}%
  \BibitemOpen
  \bibfield  {author} {\bibinfo {author} {\bibfnamefont {N.}~\bibnamefont
  {Bergmann}}\ and\ \bibinfo {author} {\bibfnamefont {M.}~\bibnamefont
  {Galperin}},\ }\href {\doibase arXiv:2004.05175} {\bibfield  {journal}
  {\bibinfo  {journal} {arXiv:2004.05175 [cond-mat]}\
  }arXiv:2004.05175}\BibitemShut {NoStop}%
\bibitem [{\citenamefont {Bruch}\ \emph {et~al.}(2016)\citenamefont {Bruch},
  \citenamefont {Thomas}, \citenamefont {Viola~Kusminskiy}, \citenamefont {{von
  Oppen}},\ and\ \citenamefont {Nitzan}}]{bruch16}%
  \BibitemOpen
  \bibfield  {author} {\bibinfo {author} {\bibfnamefont {A.}~\bibnamefont
  {Bruch}}, \bibinfo {author} {\bibfnamefont {M.}~\bibnamefont {Thomas}},
  \bibinfo {author} {\bibfnamefont {S.}~\bibnamefont {Viola~Kusminskiy}},
  \bibinfo {author} {\bibfnamefont {F.}~\bibnamefont {{von Oppen}}}, \ and\
  \bibinfo {author} {\bibfnamefont {A.}~\bibnamefont {Nitzan}},\ }\href
  {\doibase 10.1103/PhysRevB.93.115318} {\bibfield  {journal} {\bibinfo
  {journal} {Phys. Rev. B}\ }\textbf {\bibinfo {volume} {93}},\ \bibinfo
  {pages} {115318} (\bibinfo {year} {2016})}\BibitemShut {NoStop}%
\bibitem [{\citenamefont {Katz}\ and\ \citenamefont
  {Kosloff}(2016)}]{katz2016}%
  \BibitemOpen
  \bibfield  {author} {\bibinfo {author} {\bibfnamefont {G.}~\bibnamefont
  {Katz}}\ and\ \bibinfo {author} {\bibfnamefont {R.}~\bibnamefont {Kosloff}},\
  }\href {\doibase 10.3390/e18050186} {\bibfield  {journal} {\bibinfo
  {journal} {Entropy}\ }\textbf {\bibinfo {volume} {18}},\ \bibinfo {pages}
  {186} (\bibinfo {year} {2016})}\BibitemShut {NoStop}%
\bibitem [{\citenamefont {Seifert}(2016)}]{seifert2016}%
  \BibitemOpen
  \bibfield  {author} {\bibinfo {author} {\bibfnamefont {U.}~\bibnamefont
  {Seifert}},\ }\href {\doibase 10.1103/PhysRevLett.116.020601} {\bibfield
  {journal} {\bibinfo  {journal} {Phys. Rev. Lett.}\ }\textbf {\bibinfo
  {volume} {116}},\ \bibinfo {pages} {020601} (\bibinfo {year}
  {2016})}\BibitemShut {NoStop}%
\bibitem [{\citenamefont {Freitas}\ and\ \citenamefont
  {Paz}(2017)}]{freitas2017}%
  \BibitemOpen
  \bibfield  {author} {\bibinfo {author} {\bibfnamefont {N.}~\bibnamefont
  {Freitas}}\ and\ \bibinfo {author} {\bibfnamefont {J.~P.}\ \bibnamefont
  {Paz}},\ }\href {\doibase 10.1103/PhysRevE.95.012146} {\bibfield  {journal}
  {\bibinfo  {journal} {Phys. Rev. E}\ }\textbf {\bibinfo {volume} {95}},\
  \bibinfo {pages} {012146} (\bibinfo {year} {2017})}\BibitemShut {NoStop}%
\bibitem [{\citenamefont {Perarnau-Llobet}\ \emph {et~al.}(2018)\citenamefont
  {Perarnau-Llobet}, \citenamefont {Wilming}, \citenamefont {Riera},
  \citenamefont {Gallego},\ and\ \citenamefont {Eisert}}]{perarnau2018}%
  \BibitemOpen
  \bibfield  {author} {\bibinfo {author} {\bibfnamefont {M.}~\bibnamefont
  {Perarnau-Llobet}}, \bibinfo {author} {\bibfnamefont {H.}~\bibnamefont
  {Wilming}}, \bibinfo {author} {\bibfnamefont {A.}~\bibnamefont {Riera}},
  \bibinfo {author} {\bibfnamefont {R.}~\bibnamefont {Gallego}}, \ and\
  \bibinfo {author} {\bibfnamefont {J.}~\bibnamefont {Eisert}},\ }\href
  {\doibase 10.1103/PhysRevLett.120.120602} {\bibfield  {journal} {\bibinfo
  {journal} {Phys. Rev. Lett.}\ }\textbf {\bibinfo {volume} {120}},\ \bibinfo
  {pages} {120602} (\bibinfo {year} {2018})}\BibitemShut {NoStop}%
\bibitem [{\citenamefont {Bruch}\ \emph {et~al.}(2018)\citenamefont {Bruch},
  \citenamefont {Lewenkopf},\ and\ \citenamefont {{von Oppen}}}]{bruch18}%
  \BibitemOpen
  \bibfield  {author} {\bibinfo {author} {\bibfnamefont {A.}~\bibnamefont
  {Bruch}}, \bibinfo {author} {\bibfnamefont {C.}~\bibnamefont {Lewenkopf}}, \
  and\ \bibinfo {author} {\bibfnamefont {F.}~\bibnamefont {{von Oppen}}},\
  }\href {\doibase 10.1103/PhysRevLett.120.107701} {\bibfield  {journal}
  {\bibinfo  {journal} {Phys. Rev. Lett.}\ }\textbf {\bibinfo {volume} {120}},\
  \bibinfo {pages} {107701} (\bibinfo {year} {2018})}\BibitemShut {NoStop}%
\bibitem [{\citenamefont {Dou}\ \emph {et~al.}(2018)\citenamefont {Dou},
  \citenamefont {Ochoa}, \citenamefont {Nitzan},\ and\ \citenamefont
  {Subotnik}}]{dou2018}%
  \BibitemOpen
  \bibfield  {author} {\bibinfo {author} {\bibfnamefont {W.}~\bibnamefont
  {Dou}}, \bibinfo {author} {\bibfnamefont {M.~A.}\ \bibnamefont {Ochoa}},
  \bibinfo {author} {\bibfnamefont {A.}~\bibnamefont {Nitzan}}, \ and\ \bibinfo
  {author} {\bibfnamefont {J.~E.}\ \bibnamefont {Subotnik}},\ }\href {\doibase
  10.1103/PhysRevB.98.134306} {\bibfield  {journal} {\bibinfo  {journal} {Phys.
  Rev. B}\ }\textbf {\bibinfo {volume} {98}},\ \bibinfo {pages} {134306}
  (\bibinfo {year} {2018})}\BibitemShut {NoStop}%
\bibitem [{\citenamefont {Haughian}\ \emph {et~al.}(2018)\citenamefont
  {Haughian}, \citenamefont {Esposito},\ and\ \citenamefont
  {Schmidt}}]{haughian18}%
  \BibitemOpen
  \bibfield  {author} {\bibinfo {author} {\bibfnamefont {P.}~\bibnamefont
  {Haughian}}, \bibinfo {author} {\bibfnamefont {M.}~\bibnamefont {Esposito}},
  \ and\ \bibinfo {author} {\bibfnamefont {T.~L.}\ \bibnamefont {Schmidt}},\
  }\href {\doibase 10.1103/PhysRevB.97.085435} {\bibfield  {journal} {\bibinfo
  {journal} {Phys. Rev. B}\ }\textbf {\bibinfo {volume} {97}},\ \bibinfo
  {pages} {085435} (\bibinfo {year} {2018})}\BibitemShut {NoStop}%
\bibitem [{\citenamefont {Strasberg}(2019)}]{strasberg2019}%
  \BibitemOpen
  \bibfield  {author} {\bibinfo {author} {\bibfnamefont {P.}~\bibnamefont
  {Strasberg}},\ }\href {\doibase 10.1103/PhysRevE.100.022127} {\bibfield
  {journal} {\bibinfo  {journal} {Phy. Rev. E}\ }\textbf {\bibinfo {volume}
  {100}},\ \bibinfo {pages} {022127} (\bibinfo {year} {2019})}\BibitemShut
  {NoStop}%
\bibitem [{\citenamefont {Martinez}\ and\ \citenamefont
  {Paz}(2013)}]{martinez2013}%
  \BibitemOpen
  \bibfield  {author} {\bibinfo {author} {\bibfnamefont {E.~A.}\ \bibnamefont
  {Martinez}}\ and\ \bibinfo {author} {\bibfnamefont {J.~P.}\ \bibnamefont
  {Paz}},\ }\href {\doibase 10.1103/PhysRevLett.110.130406} {\bibfield
  {journal} {\bibinfo  {journal} {Phys. Rev. Lett.}\ }\textbf {\bibinfo
  {volume} {110}},\ \bibinfo {pages} {130406} (\bibinfo {year}
  {2013})}\BibitemShut {NoStop}%
\bibitem [{\citenamefont {Freitas}\ and\ \citenamefont
  {Paz}(2014)}]{freitas2014}%
  \BibitemOpen
  \bibfield  {author} {\bibinfo {author} {\bibfnamefont {N.}~\bibnamefont
  {Freitas}}\ and\ \bibinfo {author} {\bibfnamefont {J.~P.}\ \bibnamefont
  {Paz}},\ }\href {\doibase 10.1103/PhysRevE.90.042128} {\bibfield  {journal}
  {\bibinfo  {journal} {Phys. Rev. E}\ }\textbf {\bibinfo {volume} {90}},\
  \bibinfo {pages} {042128} (\bibinfo {year} {2014})}\BibitemShut {NoStop}%
\bibitem [{\citenamefont {Vool}\ and\ \citenamefont
  {Devoret}(2016)}]{devoret2016}%
  \BibitemOpen
  \bibfield  {author} {\bibinfo {author} {\bibfnamefont {U.}~\bibnamefont
  {Vool}}\ and\ \bibinfo {author} {\bibfnamefont {M.~H.}\ \bibnamefont
  {Devoret}},\ }\href {\doibase 10.1093/acprof:oso/9780199681181.003.0003}
  {\bibfield  {journal} {\bibinfo  {journal} {International Journal of Circuit
  Theory and Applications}\ }\textbf {\bibinfo {volume} {45}},\ \bibinfo
  {pages} {895} (\bibinfo {year} {2016})}\BibitemShut {NoStop}%
\bibitem [{\citenamefont {Girvin}(2014)}]{girvin2014}%
  \BibitemOpen
  \bibfield  {author} {\bibinfo {author} {\bibfnamefont {S.~M.}\ \bibnamefont
  {Girvin}},\ }\href {\doibase 10.1002/cta.2359} {\bibfield  {journal}
  {\bibinfo  {journal} {Oxford University Press}\ } (\bibinfo {year} {2014}),\
  10.1002/cta.2359}\BibitemShut {NoStop}%
\bibitem [{\citenamefont {Gaul}\ and\ \citenamefont
  {B{\"u}ttner}(2007)}]{gaul2007}%
  \BibitemOpen
  \bibfield  {author} {\bibinfo {author} {\bibfnamefont {C.}~\bibnamefont
  {Gaul}}\ and\ \bibinfo {author} {\bibfnamefont {H.}~\bibnamefont
  {B{\"u}ttner}},\ }\href {\doibase 10.1103/PhysRevE.76.011111} {\bibfield
  {journal} {\bibinfo  {journal} {Phys. Rev. E}\ }\textbf {\bibinfo {volume}
  {76}},\ \bibinfo {pages} {011111} (\bibinfo {year} {2007})}\BibitemShut
  {NoStop}%
\bibitem [{\citenamefont {Asadian}\ \emph {et~al.}(2013)\citenamefont
  {Asadian}, \citenamefont {Manzano}, \citenamefont {Tiersch},\ and\
  \citenamefont {Briegel}}]{asadian2013}%
  \BibitemOpen
  \bibfield  {author} {\bibinfo {author} {\bibfnamefont {A.}~\bibnamefont
  {Asadian}}, \bibinfo {author} {\bibfnamefont {D.}~\bibnamefont {Manzano}},
  \bibinfo {author} {\bibfnamefont {M.}~\bibnamefont {Tiersch}}, \ and\
  \bibinfo {author} {\bibfnamefont {H.}~\bibnamefont {Briegel}},\ }\href
  {\doibase 10.1103/PhysRevE.87.012109} {\bibfield  {journal} {\bibinfo
  {journal} {Phy. Rev. E}\ }\textbf {\bibinfo {volume} {87}},\ \bibinfo {pages}
  {012109} (\bibinfo {year} {2013})}\BibitemShut {NoStop}%
\bibitem [{\citenamefont {Nicacio}\ \emph {et~al.}(2015)\citenamefont
  {Nicacio}, \citenamefont {Ferraro}, \citenamefont {Imparato}, \citenamefont
  {Paternostro},\ and\ \citenamefont {Semi{\~a}o}}]{nicacio2015}%
  \BibitemOpen
  \bibfield  {author} {\bibinfo {author} {\bibfnamefont {F.}~\bibnamefont
  {Nicacio}}, \bibinfo {author} {\bibfnamefont {A.}~\bibnamefont {Ferraro}},
  \bibinfo {author} {\bibfnamefont {A.}~\bibnamefont {Imparato}}, \bibinfo
  {author} {\bibfnamefont {M.}~\bibnamefont {Paternostro}}, \ and\ \bibinfo
  {author} {\bibfnamefont {F.}~\bibnamefont {Semi{\~a}o}},\ }\href {\doibase
  10.1103/PhysRevE.91.042116} {\bibfield  {journal} {\bibinfo  {journal} {Phys.
  Rev. E}\ }\textbf {\bibinfo {volume} {91}},\ \bibinfo {pages} {042116}
  (\bibinfo {year} {2015})}\BibitemShut {NoStop}%
\bibitem [{\citenamefont {Freitas}\ \emph {et~al.}(2015)\citenamefont
  {Freitas}, \citenamefont {Martinez},\ and\ \citenamefont
  {Paz}}]{freitas2015}%
  \BibitemOpen
  \bibfield  {author} {\bibinfo {author} {\bibfnamefont {N.}~\bibnamefont
  {Freitas}}, \bibinfo {author} {\bibfnamefont {E.~A.}\ \bibnamefont
  {Martinez}}, \ and\ \bibinfo {author} {\bibfnamefont {J.~P.}\ \bibnamefont
  {Paz}},\ }\href {\doibase 10.1088/0031-8949/91/1/013007} {\bibfield
  {journal} {\bibinfo  {journal} {Physica Scripta}\ }\textbf {\bibinfo {volume}
  {91}},\ \bibinfo {pages} {013007} (\bibinfo {year} {2015})}\BibitemShut
  {NoStop}%
\bibitem [{\citenamefont {Riseborough}\ \emph {et~al.}(1985)\citenamefont
  {Riseborough}, \citenamefont {Hanggi},\ and\ \citenamefont
  {Weiss}}]{riseborough1985}%
  \BibitemOpen
  \bibfield  {author} {\bibinfo {author} {\bibfnamefont {P.~S.}\ \bibnamefont
  {Riseborough}}, \bibinfo {author} {\bibfnamefont {P.}~\bibnamefont {Hanggi}},
  \ and\ \bibinfo {author} {\bibfnamefont {U.}~\bibnamefont {Weiss}},\ }\href
  {\doibase 10.1103/PhysRevA.31.471} {\bibfield  {journal} {\bibinfo  {journal}
  {Phys. Rev. A}\ }\textbf {\bibinfo {volume} {31}},\ \bibinfo {pages} {471}
  (\bibinfo {year} {1985})}\BibitemShut {NoStop}%
\bibitem [{\citenamefont {Mascherpa}\ \emph {et~al.}(2020)\citenamefont
  {Mascherpa}, \citenamefont {Smirne}, \citenamefont {Somoza}, \citenamefont
  {Fern{\ifmmode\acute{a}\else\'{a}\fi}ndez-Acebal}, \citenamefont {Donadi},
  \citenamefont {Tamascelli}, \citenamefont {Huelga},\ and\ \citenamefont
  {Plenio}}]{Mascherpa2020}%
  \BibitemOpen
  \bibfield  {author} {\bibinfo {author} {\bibfnamefont {F.}~\bibnamefont
  {Mascherpa}}, \bibinfo {author} {\bibfnamefont {A.}~\bibnamefont {Smirne}},
  \bibinfo {author} {\bibfnamefont {A.~D.}\ \bibnamefont {Somoza}}, \bibinfo
  {author} {\bibfnamefont {P.}~\bibnamefont
  {Fern{\ifmmode\acute{a}\else\'{a}\fi}ndez-Acebal}}, \bibinfo {author}
  {\bibfnamefont {S.}~\bibnamefont {Donadi}}, \bibinfo {author} {\bibfnamefont
  {D.}~\bibnamefont {Tamascelli}}, \bibinfo {author} {\bibfnamefont {S.~F.}\
  \bibnamefont {Huelga}}, \ and\ \bibinfo {author} {\bibfnamefont {M.~B.}\
  \bibnamefont {Plenio}},\ }\href {\doibase 10.1103/PhysRevA.101.052108}
  {\bibfield  {journal} {\bibinfo  {journal} {Phys. Rev. A}\ }\textbf {\bibinfo
  {volume} {101}},\ \bibinfo {pages} {052108} (\bibinfo {year}
  {2020})}\BibitemShut {NoStop}%
\bibitem [{\citenamefont {Pascal}\ \emph {et~al.}(2011)\citenamefont {Pascal},
  \citenamefont {Courtois},\ and\ \citenamefont {Hekking}}]{Pascal2011}%
  \BibitemOpen
  \bibfield  {author} {\bibinfo {author} {\bibfnamefont {L.~M.~A.}\
  \bibnamefont {Pascal}}, \bibinfo {author} {\bibfnamefont {H.}~\bibnamefont
  {Courtois}}, \ and\ \bibinfo {author} {\bibfnamefont {F.~W.~J.}\ \bibnamefont
  {Hekking}},\ }\href {\doibase 10.1103/PhysRevB.83.125113} {\bibfield
  {journal} {\bibinfo  {journal} {Phys. Rev. B}\ }\textbf {\bibinfo {volume}
  {83}},\ \bibinfo {pages} {125113} (\bibinfo {year} {2011})}\BibitemShut
  {NoStop}%
\bibitem [{\citenamefont {Pekola}(2015)}]{Pekola2015}%
  \BibitemOpen
  \bibfield  {author} {\bibinfo {author} {\bibfnamefont {J.~P.}\ \bibnamefont
  {Pekola}},\ }\href {\doibase 10.1038/nphys3169} {\bibfield  {journal}
  {\bibinfo  {journal} {Nat. Phys.}\ }\textbf {\bibinfo {volume} {11}},\
  \bibinfo {pages} {118} (\bibinfo {year} {2015})}\BibitemShut {NoStop}%
\bibitem [{\citenamefont {Ronzani}\ \emph {et~al.}(2018)\citenamefont
  {Ronzani}, \citenamefont {Karimi}, \citenamefont {Senior}, \citenamefont
  {Chang}, \citenamefont {Peltonen}, \citenamefont {Chen},\ and\ \citenamefont
  {Pekola}}]{Ronzani2018}%
  \BibitemOpen
  \bibfield  {author} {\bibinfo {author} {\bibfnamefont {A.}~\bibnamefont
  {Ronzani}}, \bibinfo {author} {\bibfnamefont {B.}~\bibnamefont {Karimi}},
  \bibinfo {author} {\bibfnamefont {J.}~\bibnamefont {Senior}}, \bibinfo
  {author} {\bibfnamefont {Y.-C.}\ \bibnamefont {Chang}}, \bibinfo {author}
  {\bibfnamefont {J.~T.}\ \bibnamefont {Peltonen}}, \bibinfo {author}
  {\bibfnamefont {C.}~\bibnamefont {Chen}}, \ and\ \bibinfo {author}
  {\bibfnamefont {J.~P.}\ \bibnamefont {Pekola}},\ }\href {\doibase
  10.1038/s41567-018-0199-4} {\bibfield  {journal} {\bibinfo  {journal} {Nat.
  Phys.}\ }\textbf {\bibinfo {volume} {14}},\ \bibinfo {pages} {991} (\bibinfo
  {year} {2018})}\BibitemShut {NoStop}%
\bibitem [{\citenamefont {Li}\ \emph {et~al.}(2019)\citenamefont {Li},
  \citenamefont
  {Fern{\ifmmode\acute{a}\else\'{a}\fi}ndez-Alc{\ifmmode\acute{a}\else\'{a}\fi}zar},
  \citenamefont {Ellis}, \citenamefont {Shapiro},\ and\ \citenamefont
  {Kottos}}]{Li2019}%
  \BibitemOpen
  \bibfield  {author} {\bibinfo {author} {\bibfnamefont {H.}~\bibnamefont
  {Li}}, \bibinfo {author} {\bibfnamefont {L.~J.}\ \bibnamefont
  {Fern{\ifmmode\acute{a}\else\'{a}\fi}ndez-Alc{\ifmmode\acute{a}\else\'{a}\fi}zar}},
  \bibinfo {author} {\bibfnamefont {F.}~\bibnamefont {Ellis}}, \bibinfo
  {author} {\bibfnamefont {B.}~\bibnamefont {Shapiro}}, \ and\ \bibinfo
  {author} {\bibfnamefont {T.}~\bibnamefont {Kottos}},\ }\href {\doibase
  10.1103/PhysRevLett.123.165901} {\bibfield  {journal} {\bibinfo  {journal}
  {Phys. Rev. Lett.}\ }\textbf {\bibinfo {volume} {123}},\ \bibinfo {pages}
  {165901} (\bibinfo {year} {2019})}\BibitemShut {NoStop}%
\end{thebibliography}%

\appendix
\section{Coupled RLC circuits Hamiltonian}
\label{ap:hamiltonian}
The Hamiltonian of a LC circuit can be  written such that
	\begin{equation}\label{ap-eq:LC_classic_hamiltonian}
	H=\frac{\phi^2}{2L}+\frac{q^2}{2C}.
	\end{equation}
In this notation, $q$ and $\phi$ will play the role of momentum and position conjugate variables, respectively. To quantize the LC circuit, we need to replace the classical variables of the Hamiltonian~\eqref{ap-eq:LC_classic_hamiltonian} with their quantum counterparts. In the other words, the Poisson bracket of the flux and charge in the circuit would be
	\begin{equation}
	\lbrace\phi,q\rbrace=\frac{\partial\phi}{\partial\phi}\frac{\partial q}{\partial q}-\frac{\partial q}{\partial\phi}\frac{\partial\phi}{\partial q}=1.
	\end{equation}
As shown by Dirac the value of a classical Poisson bracket imposes it's corresponding quantum commutator
	\begin{equation}
	\lbrace \phi,q\rbrace \rightarrow \frac{1}{i\hbar}[\hat{\phi},\hat{q}].
	\end{equation}
Thus, we see that transforming the classical Hamiltonian into its quantum version will also be backed by the uncertainty relation between flux and charge variables as they play the role of position and momentum, respectively.

\begin{figure}[t]
\centering
	\includegraphics[width=0.8\linewidth]{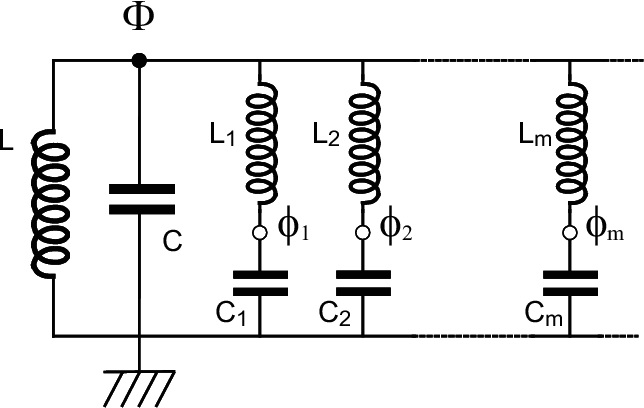}
	\caption{A damped LC circuit.}\label{ap-fig:damped_LC_circuit}
\end{figure}

 The dissipative part would be the resistor attached to the LC circuit. However, adding the Hamiltonian of this part is not trivial. To write the full Hamiltonian of a RLC circuit, we will employ the Caldeira-Leggett model for the Brownian motion. The resistor can be considered as a circuit consisting of an infinite array of independent LC circuits each playing the role of harmonic oscillators of the bath (Fig.~\ref{ap-fig:damped_LC_circuit}).
Considering the RLC circuit of Fig.~\ref{ap-fig:damped_LC_circuit}, we can write the full Hamiltonian describing the circuit such that
	\begin{equation}\label{ap-eq:damped_LC_hamiltonian}
	H=\frac{\Phi^2}{2L}+\frac{q^2}{2C}+\sum_m \frac{q_{m}^2}{2C_m}+\frac{(\phi_{m}-\Phi)^2}{2L_{m}}.
	\end{equation}
In the above expression, the flux variable $\Phi$ and $\phi_{m}$ correspond to the \textit{node} fluxes. The node flux is defined as the time integral of the voltage along the path connecting the node and the ground. $q$ and $q_m$ are also the charge in the capacitor $C$ and $C_m$.  The last term in the Hamiltonian can also be realized as the normalizing term to ensure that there will be no inconsistency in the minimum of the potential energy.

\begin{figure}[t]
  \centering
  \includegraphics[scale=.85]{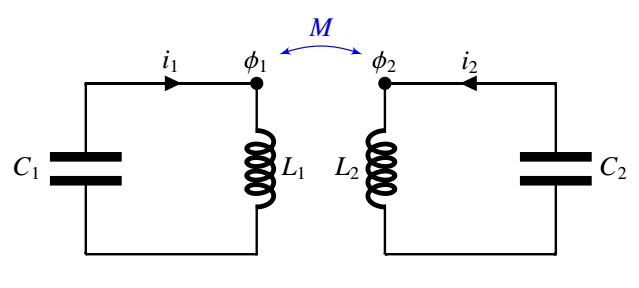}
    \caption{Two magnetically coupled circuits.}\label{ap-fig:simple_coupled_circuits}
  \end{figure}

Next, we will magnetically couple two quantum RLC circuits by putting them into the proximity of each other. Indeed, the coupling occurs due to the presence of a flux running in one circuit which is caused by the other inductor. This leads to the mutual inductance between the two inductors of the two circuits.
Before considering two coupled RLC circuits, we first look at two simple coupled circuits of Fig.~\ref{ap-fig:simple_coupled_circuits}. We denote the total flux passing through $l$-th circuit by $\phi_{l}$ with $l=\{1,2\}$. The total flux is the sum over the flux $\phi_{ll}$ produced by the inductor $L_{l}$ and the mutual flux $\phi_{lk}$ between the circuits with $k=\{1,2\}$ . We may write this such that
	\begin{align}\label{ap-eq:phase_relation}
	\phi_{l}=\sum_{k}\phi_{lk}.
	\end{align}
To find the Hamiltonian of the coupled circuits, we use the Kirchhoff's law for voltages to obtain
	\begin{align}\label{ap-eq: motion_coupled}
	v_{1}=&\dot{\phi}_{1}=L_{1}\frac{di_{1}}{dt}+\dot{\phi}_{12}\\
	v_{2}=&\dot{\phi}_{2}=L_{2}\frac{di_{2}}{dt}+\dot{\phi}_{21}.
	\end{align}
Where $v_{1}$ and $v_{2}$ are the voltages associated with the two capacitors and $i_{1}$ and $i_{2}$ are the currents for each circuits. For mutual flux we can write
	\begin{align}
	\dot{\phi}_{12}=&\frac{d\phi_{12}}{di_{2}}\frac{di_{2}}{dt}=M_{12}\frac{di_{2}}{dt}\nonumber\\
	\dot{\phi}_{21}=&\frac{d\phi_{21}}{di_{1}}\frac{di_{1}}{dt}=M_{21}\frac{di_{1}}{dt},
	\end{align}
where $M_{12/21}=\frac{d\phi_{12/21}}{di_{2/1}}$ is the mutual inductance between the two circuits and it can be proved that $M_{12}=M_{21}=M$.
To calculate the energy stored in the two coupled circuits, we first assume that $i_{2}=0$ and $i_{1}$ is increased up to an arbitrary value $I_{1}$. Then the power stored in the left circuit is
	\begin{equation}
	p_{1}=v_{1}i_{1}=i_{1}L_{1}\frac{d i_1}{dt}.
	\end{equation}
Then the total energy will be
	\begin{equation}
	E_{1}=\int p_{1}dt=\int_{0}^{I_{1}}i_{1}di_{1}=\frac{1}{2}L_{1}I_{1}^{2}.
	\end{equation}
Now, we assume that $i_{1}=I_{1}$ is constant and we change $i_{2}$ from zero to $I_{2}$. Since $i_{2}$ is changing, the mutual voltage induced in the left circuit is $M di_{2}/dt$ and therefore the total power will become
	\begin{equation}
	p_{2}=v_{2}i_{2}+I_{1}M\frac{di_{2}}{dt}=i_{2}L_{2}\frac{d i_2}{dt}+I_{1}M\frac{di_{2}}{dt},
	\end{equation}
thus the energy stored in the circuit can be written as
	\begin{equation}
	E_{2}\!=\!\!\int\!\! p_{2}dt\!=\!L_2\int_{0}^{I_{2}}\!\!i_{2}di_{2}+I_1\int_{0}^{I_{2}}\!\!Mdi_{2}\!=\frac{1}{2}L_{2}I_{2}^{2}+MI_{1}I_{2}.
	\end{equation}
We can write the total energy of the circuits as the sum over $E_1$ and $E_2$ such that
	\begin{equation}
	E_{1}+E_{2}=\frac{1}{2}L_{1}I_{1}^{2}+\frac{1}{2}L_{2}I_{2}^{2}+MI_{1}I_{2}.
\end{equation}
Adding the energy with respect to the capacitors to this energy we can write the Hamiltonian as
	\begin{equation}\label{ap-eq:couplde_circuits_energy}
	H=\frac{q_{1}^2}{2C_{1}}+\frac{q_{2}^2}{2C_{2}}+\frac{1}{2}L_{1}i_{1}^{2}+\frac{1}{2}L_{2}i_{2}^{2}+Mi_{1}i_{2}.
	\end{equation}
Above, we replaced arbitrary currents $I_{1}$ and $I_{2}$ by $i_{1}$ and $i_{2}$. We can see that $M$ is the coupling constant between the two circuits.

To find the Hamiltonian of the two RLC circuits, similar to what we did in Eq.~\eqref{ap-eq:damped_LC_hamiltonian} we attach two resistors to the both ends of the coupled LC circuits. We can replace currents in Eq.~\eqref{ap-eq:couplde_circuits_energy} with their flux variables by using the relation $\phi_{ll}=L_li_l$. Doing so the Hamiltonian of this model will then become

	\begin{align}\label{ap-eq:model_hamiltonian_self_flux}
	 H&=
\frac{q_{1}^{2}}{2C_{1}}\!+\!\frac{q_{2}^{2}}{2C_{2}}
\!+\!\frac{\phi_{11}^{2}}{2L_{1}}\!+\frac{\phi_{22}^{2}}{2L_{2}}
\!+\!\frac{M}{L_{1}L_{2}}\phi_{11}\phi_{22}\nonumber\\
	&+\!\sum_{m_{1}}\frac{q_{m_{1}}^{2}}{2C_{m_{1}}}\!+\!\frac{(\phi_{m_{1}}-\phi_{1})^{2}}{2L_{m_{1}}}\!+\!\sum_{m_2}\!\frac{q_{m_{2}}^{2}}{2C_{m_{2}}}\!+\!\frac{(\phi_{m_{2}}-\phi_{2})^{2}}{2L_{m_{2}}}.
	\end{align}
Next, we will eliminate the flux terms $\phi_{ll}$ to write it in terms of the total flux $\phi_l$. To do so we use the bellow relations between the fluxes
	\begin{align}
	\phi_{12}=&M_{12}i_{2}=\frac{M}{L_{2}}\phi_{22}\\
	\phi_{21}=&M_{21}i_{1}=\frac{M}{L_{1}}\phi_{11}
		\end{align}
	These relations together with Eq.~\eqref{ap-eq:phase_relation} gives
		\begin{align}
	\phi_{11}=&\frac{L_1L_2}{M^{2}-L_{1}L_{2}}\left(\frac{M}{L_{2}}\phi_{2}-\phi_{1}\right)\nonumber\\
	\phi_{22}=&\frac{L_1L_2}{M^{2}-L_{1}L_{2}}\left(\frac{M}{L_{1}}\phi_{1}-\phi_{2}\right).
	\end{align}
We can now replace these transformation in Eq.~\eqref{ap-eq:model_hamiltonian_self_flux} to find the Hamiltonian of our model such that
\begin{align}
H&=
\frac{q_{1}^{2}}{2C_{1}}\!+\!\frac{q_{2}^{2}}{2C_{2}}
\!+\!\frac{L_1L_2}{L_{1}L_{2}\!-\!M^{2}}
\!\left[\frac{\phi_{1}^{2}}{2L_{1}}\!+\!
\frac{\phi_{2}^{2}}{2L_{2}}\!-\!
\frac{M}{L_{1}L_{2}}\phi_{1}\phi_{2}\right]\nonumber\\
&+\!\sum_{m_{1}}\!
\left(\!\frac{q_{m_{1}}^{2}}{2C_{m_{1}}}\!+\!
\frac{(\phi_{m_{1}}\!-\!\phi_{1})^{2}}{2L_{m_{1}}}\!\right)\!+\!
\!\sum_{m_{2}}\!
\left(\!\frac{q_{m_{2}}^{2}}{2C_{m_{2}}}\!+\!
\frac{\!(\phi_{m_{2}}\!-\!\phi_{2})^{2}}{2L_{m_{2}}}\!\right).
	\end{align}

\section{Steady state heat currents}\label{ap:heat_current_proof}
Using the Heisenberg equations of motion, we can find the integro-differential equation for each variables of the system and then they can be solved using the Green's function matrix of the system $g(t,t^{\prime})$ which satisfies the integro-differential equation,
	\begin{equation}\label{ap-eq:system_green_function }
	C\frac{\partial^{2}}{\partial t^{2}}g(t,t^{\prime})+L^{-1}_0 g(t,t^{\prime})+\!\int_{0}^{t}\!\!\!\gamma(t-\tau)\frac{\partial}{\partial \tau} g(\tau,t^{\prime})d\tau\!=\!\delta(t-t^\prime),
	\end{equation}
with the initial condition $g(0,t^{\prime})=0$.

Assuming that the Green's function $g(t,t^{\prime})$ is an exponentially decaying function with respect to $t$, the correlations functions between the system variables, i.e $\phi$ and $q$, will be independent of the initial state of the total system for large $t$. To capture the correlation functions, we use covariance matrix $\sigma$ such that
 \begin{equation}
 \sigma=
 \begin{bmatrix}
 \sigma^{(\phi,\phi)} && \sigma^{(\phi,q)}\\
 \sigma^{(q,\phi)} && \sigma^{(q,q)}
 \end{bmatrix}.
 \end{equation}
 In terms of the Green's function $g(t,t^{\prime})$, one can obtain
\begin{align}\label{ap-eq:covariance_components}
 	\sigma^{(n,m)}(t)\!=&\frac{\hbar}{2}\int_{0}^{t}\!\!\int_{0}^{t}\!\!g^{(n)}(t,t_{1})\nu_{\alpha}(t_{1}-t_{2})g^{(m)}(t,t_{2})^{T}dt_{1}dt_{2},
 	\end{align}
 where
\begin{equation}
\nu_{\alpha}(t)=\int_{0}^{\infty}\!\!d\omega I_{\alpha}(\omega)\cos(\omega t)\coth\left(\frac{\hbar\beta_{\alpha}\omega}{2}\right),
\end{equation}
denotes the noise kernel. Also $\sigma^{(0,0)}=\sigma^{(\phi,\phi)}$, $\sigma^{(0,1)}=\sigma^{(\phi,q)}$ and $\sigma^{(1,1)}=\sigma^{(q,
q)}$ and $g^{(n)}$ is the $n$th derivative of $g$. Considering a situation in which, the spectral density is a continuous function of $\omega$  we can write the covariance matrix for the steady state limit, i.e $t\rightarrow\infty$ such that
	\begin{align}\label{ap-eq:steady_state_covariance}
	\sigma^{(n,m)}=&\mathrm{Re}\int_{0}^{\infty}\frac{\hbar}{2}\omega^{n+m}i^{n-m}g(i\omega)\nu_{\alpha}(\omega)g(-i\omega)^{T}C d\omega.
	\end{align}
Where $\sigma^{(n,m)}$ is the covariance matrix in the asymptotic state, $\nu_{\alpha}(\omega)$ is the Fourier transform  of the noise kernel and $g(s)$ is the Laplace transform of the Green function which can be obtained using Eq.~\eqref{ap-eq:system_green_function } such that
 	\begin{equation}\label{ap-eq:green_laplace}
	g(s)^{-1}=Cs^2+\gamma(s)s+L^{-1}_{0},
	\end{equation}
where $\gamma(s)$ is the Laplace transform of $\gamma(t)$.

Now we turn to analyse the heat flow thorough the system. Since there exists no deriving in the system, the heat current is directly related to the change in the mean value of the energy of the system and for the steady state one can write
\begin{align}\label{ap-eq:heat_current_trace_only}
	\dot{Q}_{\alpha}=\mathrm{Tr}\left[P_{\alpha}L^{-1}\sigma^{(\phi, q)}(t)C^{-1}\right].
	\end{align}
 To calculate the local heat current, we first write $\sigma^{(\phi, q)}(t)$ by using
	\begin{equation}
	\nu_{\alpha}(t_1-t_2)=\mathrm{Re}\int_{0}^{\infty}\nu_{\alpha}(\omega)e^{-i\omega(t_1-t_2)}d\omega,
	\end{equation}
where $\nu_{\alpha}(\omega)=I_{\alpha}(\omega)\coth(\frac{\hbar \beta_{\alpha}\omega}{2})$. Replacing this into Eq.~\eqref{ap-eq:covariance_components} we can see in the limit $t\rightarrow\infty$ we can write
	\begin{align}
	\int_{0}^{\infty}g(t,t_{1})e^{-i\omega t_{1}}dt_{1}=g(i\omega),
	\end{align}
thus we will have
	\begin{align}
	\sigma^{\phi q}(t)=&-\mathrm{Re}\int_{0}^{\infty}\frac{\hbar}{2}g(i\omega)\nu_{\alpha}(\omega)i\omega g(-i\omega)^{T}C d\omega.
	\end{align}
Replacing this equation into Eq.~\eqref{ap-eq:heat_current_trace_only} we will have the local heat current expression for steady state limit as
	\begin{align}\label{ap-eq:heat_current_all_baths}
	\dot{Q}_{\alpha}=\frac{\hbar}{2}\sum_{\alpha^{\prime}}\int_{0}^{\infty}\omega f_{\alpha\alpha^{\prime}}(\omega)\coth(\frac{\hbar \beta_{\alpha^{\prime}}\omega}{2})d\omega.
	\end{align}
Where we have used the fact that $\mathrm{Re}(-iX)=\mathrm{Im}(X)$. The heat transfer matrix element $f_{\alpha\alpha^{\prime}}$ is written such that
	\begin{equation}
	f_{\alpha\alpha^{\prime}}(\omega)=\mathrm{Im}\mathrm{Tr}\left[P_{\alpha}L^{-1}g(i\omega)I_{\alpha^{\prime}}(\omega) g(-i\omega)^{T}\right]
	\end{equation}
Here we have $P_{\alpha}L^{-1}=P_{\alpha}L^{-1}_{0}+L^{-1}_{\alpha}$. Replacing this relation into the above equation, we can see that $ \mathrm{Tr}\left[P_{\alpha}L_{\alpha}^{-1}g(i\omega)I_{\alpha^{\prime}}(\omega) g(-i\omega)^{T}\right]=0$ because $L^{-1}_{\alpha}$ is a symmetric matrix and $g(i\omega)I_{\alpha^{\prime}}(\omega) g(-i\omega)^{T}$ is anti-symmetric and the trace of their product will be vanishing. 
 That said we can write the heat transfer matrix as
	\begin{equation}\label{ap-eq:heat_transfer}
	f_{\alpha\alpha^{\prime}}(\omega)=\mathrm{Im}\mathrm{Tr}\left[P_{\alpha}L^{-1}_{0}g(i\omega)I_{\alpha^{\prime}}(\omega) g(i\omega)^{\dagger}\right].
	\end{equation}
To expand the above relation a bit further, we first take the Laplace transform of Eq.~\eqref{ap-eq:system_green_function } such that
	\begin{equation}\label{green laplace}
	g(s)^{-1}=Cs^2+\gamma(s)s+L^{-1}_{0}.
	\end{equation}
Writing $L^{-1}_{0}$ in terms of $g(s)^{-1}$ we have
	\begin{equation}
	L^{-1}_{0}=g(s)^{-1}-Cs^2+\gamma(s)s.
	\end{equation}
Replacing the above equation into Eq.~\eqref{ap-eq:heat_transfer} with $s=i\omega$, we will have
	\begin{align}
	f_{\alpha\alpha^{\prime}}(\omega)=
	&\mathrm{Im}\mathrm{Tr}\left[P_{\alpha}I_{\alpha^{\prime}}(\omega) g(i\omega)^{\dagger}\right]\nonumber\\
	+&\mathrm{Im}\omega^{2}\mathrm{Tr}\left[CP_{\alpha}g(i\omega)I_{\alpha^{\prime}}(\omega) g(i\omega)^{\dagger}\right]\nonumber\\
	+&\mathrm{Im}i\omega\mathrm{Tr}\left[P_{\alpha}\gamma(i\omega)g(i\omega)I_{\alpha^{\prime}}(\omega) g(i\omega)^{\dagger}\right].
	\end{align}
The first term vanishes because $P_{\alpha}I_{\alpha^{\prime}}(\omega)=0$ for $\alpha\neq\alpha^{\prime}$. The second will also be vanishing because it is a product of two symmetric and anti-symmetric matrices. In the third term, the matrix $g(i\omega)I_{\alpha^{\prime}}(\omega) g(i\omega)^{\dagger}$ is hermitian so that we only have to calculate $\mathrm{Im}(i\omega\gamma(i\omega))=\mathrm{Re}(\omega\gamma(i\omega))=\frac{\pi}{2}I(\omega)$. Thus we have
	\begin{equation}
	f_{\alpha\alpha^{\prime}}(\omega)=\frac{\pi}{2}\mathrm{Tr}\left[I_{\alpha}(\omega)g(i\omega)I_{\alpha^{\prime}}(\omega) g(i\omega)^{\dagger}\right].
	\end{equation}
Inserting this matrix back into Eq.~\eqref{ap-eq:heat_current_all_baths} we will have
	\begin{equation}\label{ap-eq:heat_current_final}
	\dot{Q}_{\alpha}=\frac{\hbar}{2}\sum_{\alpha^{\prime}\neq\alpha}\int_{0}^{\infty}\omega d\omega f_{\alpha\alpha^{\prime}}(\omega)\left[\coth\left(\frac{\hbar \beta_{\alpha}\omega}{2}\right)-\coth\left(\frac{\hbar \beta_{\alpha^{\prime}}\omega}{2}\right)\right].
	\end{equation}
	\section{Quantum correction to the heat current}
  \label{ap:quantum_corrections}
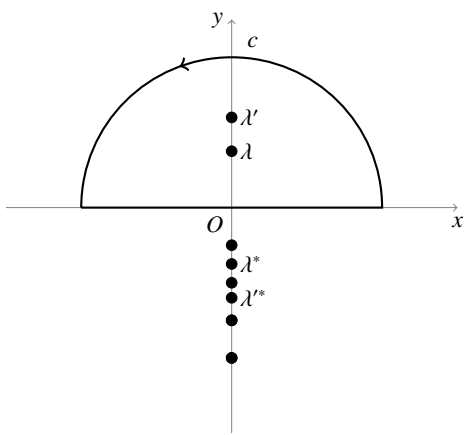
\begin{figure}[t]
\centering
\begin{tikzpicture}[decoration={markings,
mark=at position 7.85cm with {\arrow[line width=1pt]{>}},
}
]
\draw[help lines,->] (-3,0) -- (3,0) coordinate (xaxis);
\draw[help lines,->] (0,-3) -- (0,2.5) coordinate (yaxis);
\filldraw(0,-.75) circle(2pt) node[right]{$\lambda^{*}$};
\filldraw(0,-1.2) circle(2pt) node[right]{$\lambda^{\prime*}$};
\filldraw(0,-1.5) circle(2pt);
\filldraw(0,-2) circle(2pt);
\filldraw(0,-.5) circle(2pt);
\filldraw(0,-1) circle(2pt);
\filldraw(0,-1.5) circle(2pt);
\filldraw(0,-2) circle(2pt);

\filldraw(0,.75) circle(2pt) node[right]{$\lambda$};
\filldraw(0,1.2) circle(2pt) node[right]{$\lambda^{\prime}$};
\path[draw,line width=0.8pt,postaction=decorate](-2,0)--(2,0)arc(0:180:2);

\node[right] at (0.1,2.2){$c$} ;
\node[below] at (xaxis) {$x$};
\node[left] at (yaxis) {$y$};
\node[below left] {$O$};
\end{tikzpicture}
\caption{Upper-half plane contour. The dots are the poles of the digamma function.}\label{ap-fig:contour}
\end{figure}

To calculate the quantum contribution to the heat currents we will analytically solve the integral in Eq.~\eqref{eq:quantum_heat}. To do so, we will have to take into account the poles of the digamma function in addition to the poles of $f_{12}(\omega)$. In fact the poles of the function $\psi(1-ix)$ are all located on the lower-half of the imaginary axis, i.e $x=-i, -2i,-3i,....$. The poles of the heat transfer matrix element, $\lambda_{\pm}$ and their conjugates $\lambda_{\pm}^{*}$ are on the imaginary axis. However, since we would like to exclude the contribution from the digamma function poles, we choose the integration contour to run on the upper-half plane which only covers $\lambda_{\pm}$. Thus, we can write the integral such that
\begin{align}
\dot{Q}_{1}^\text{q}&=
\frac{i\hbar}{2} \! \int_{c} \! d\omega \: \omega\:
f_{12}(\omega)\left[\psi\left(\!1\!-\frac{i\beta_{2}\hbar \omega}{2\pi}\right)
\!-\psi\left(\!1\!-\frac{i\beta_{1}\hbar \omega}{2\pi}\right)\right]\nonumber\\
&+\frac{i\hbar}{2} \! \int_{\infty} \! d\omega \: \omega\:
f_{12}(\omega)\left[\psi\left(\!1\!-\frac{i\beta_{2}\hbar \omega}{2\pi}\right)
\!-\psi\left(\!1\!-\frac{i\beta_{1}\hbar \omega}{2\pi}\right)\right].
\label{ap-eq:quantum_heat}
\end{align}
The first integral is done over the contour c in Fig.~\ref{ap-fig:contour} by using the residue theorem. The second integral is the contribution for $\omega\rightarrow\infty$. In this limit we need to expand the digamma function using
	\begin{equation}\label{ap-eq:digamma_infinity}
	\psi\left(1\pm ix\right)\simeq \log\left(\pm\ ix\right)\mp \frac{i}{x},
	\end{equation}
for $x\rightarrow\infty$. Since the integrand is vanishing as $1/\omega$ then we only need to keep the logarithmic term in the asymptotic digamma functions. Replacing this expansion into the second integral in Eq.~\eqref{ap-eq:quantum_heat} we will have
\begin{align}
&\frac{i\hbar}{2}\int_{\infty} \! d\omega \: \omega\:
f_{12}(\omega)\left[\psi\left(\!1\!-\frac{i\beta_{2}\hbar \omega}{2\pi}\right)
\!-\psi\left(\!1-\frac{i\beta_{1}\hbar \omega}{2\pi}\right)\right]\nonumber\\
&=-\frac{i\hbar}{\pi}
\left(\frac{M}{L}\right)^2
\left(\frac{\lambda_+\lambda_-}{\omega_d}\right)^2
\int\!d\omega\frac{1}{\omega}\log\left(\frac{\beta_1}{\beta_2}\right).
\end{align}
We change the variable $\omega=\Lambda e^{i\theta}$ and we integrate over the semi-circle on the upper-half plane for $0\leq\theta\leq\pi$ and $\Lambda\rightarrow\infty$, thus we will have
\begin{align}
&\frac{i\hbar}{2}\int_{\infty}  d\omega \: \omega\:
f_{12}(\omega)\left[\psi\left(1-\frac{i\beta_{2}\hbar \omega}{2\pi}\right)
-\psi\left(1-\frac{i\beta_{1}\hbar \omega}{2\pi}\right)\right]\nonumber\\
&=\frac{\hbar}{\pi}
\left(\frac{M}{L}\right)^2
\left(\frac{\lambda_+\lambda_-}{\omega_d}\right)^2
\log\left(\frac{T_2}{T_1}\right).
\end{align}
Hence, by adding the above result and the integral over the contour we will obtain Eq.~\eqref{eq:quantum_heat}.

\end{document}